\documentstyle[preprint,epsf,aps,floats,tighten]{revtex}
\def\D0{D\O}                            
\def\met{\mbox{${\hbox{$E$\kern-0.6em\lower-.1ex\hbox{/}}}_T$ }} 
\def\metx{\mbox{${\hbox{$E$\kern-0.6em\lower-.1ex\hbox{/}}}_{x}$ }} 
\def\mety{\mbox{${\hbox{$E$\kern-0.6em\lower-.1ex\hbox{/}}}_{y}$ }} 
\def\d0draft{}

\def\mathunit#1{\mathop{\hbox{#1}}\mathclose{}\mathord{}}

\def\gev{\mathunit{GeV}}
\def\mev{\mathunit{MeV}}

\def\gevc{\gev \kern -1.7pt/ \kern -1.7pt c}

\def\mevc{\mev \kern -1.7pt/ \kern -1.7pt c}

\def\({\left(}
\def\){\right)}

\def\err#1#2#3 {{\it Erratum} {\bf#1},{\ #2} (19#3)}
\def\ib#1#2#3 {{\it ibid.} {\bf#1},{\ #2} (19#3)}
\def\nc#1#2#3 {Nuovo Cim. {\bf#1} ,#2(19#3)}
\def\nim#1#2#3 {Nucl. Instr. Meth. {\bf#1},{\ #2} (19#3)}
\def\np#1#2#3 {Nucl. Phys. {\bf#1},{\ #2} (19#3)}
\def\pl#1#2#3 {Phys. Lett. {\bf#1},{\ #2} (19#3)}
\def\prev#1#2#3 {Phys. Rev. {\bf#1},{\ #2} (19#3)}
\def\prl#1#2#3 {Phys. Rev. Lett. {\bf#1},{\ #2} (19#3)}
\def\rmp#1#2#3 {Rev. Mod. Phys. {\bf#1},{\ #2} (19#3)}
\def\zp#1#2#3 {Zeit. Phys. {\bf#1},{\ #2} (19#3)}
\begin{document}
\renewcommand{\textfraction}{0.0}
\renewcommand{\topfraction}{1.0}
\renewcommand{\bottomfraction}{1.0}

\title{
\vbox to 0pt{
\normalsize
\vspace{-0.5in}
\rightline{hep-ex/9912065}
\rightline{FERMILAB-Pub-99/373-E}
\vfil}
A Measurement of the {\mbox{$W \rightarrow \tau \nu$}} Production Cross Section in {\mbox{$p\overline{p}$}} Collisions at 
$\boldmath\sqrt{s}=1.8$ TeV }

%
\author{                                                                      
B.~Abbott,$^{47}$                                                             
M.~Abolins,$^{44}$                                                            
V.~Abramov,$^{19}$                                                            
B.S.~Acharya,$^{13}$                                                          
D.L.~Adams,$^{54}$                                                            
M.~Adams,$^{30}$                                                              
S.~Ahn,$^{29}$                                                                
V.~Akimov,$^{17}$                                                             
G.A.~Alves,$^{2}$                                                             
N.~Amos,$^{43}$
% ...
\\                                                                            
\vskip 0.30cm                                                                 
\centerline{(D\O\ Collaboration)}                                             
\vskip 0.30cm                                                                 
}                                                                             
\address{                                                                     
\centerline{$^{1}$Universidad de Buenos Aires, Buenos Aires, Argentina}       
\centerline{$^{2}$LAFEX, Centro Brasileiro de Pesquisas F{\'\i}sicas,         
                  Rio de Janeiro, Brazil}                                     
\centerline{$^{3}$Universidade do Estado do Rio de Janeiro,                   
                  Rio de Janeiro, Brazil}                                     
\centerline{$^{4}$Institute of High Energy Physics, Beijing,                  
                  People's Republic of China}                                 
\centerline{$^{5}$Universidad de los Andes, Bogot\'{a}, Colombia}             
\centerline{$^{6}$Universidad San Francisco de Quito, Quito, Ecuador}         
\centerline{$^{7}$Institut des Sciences Nucl\'eaires, IN2P3-CNRS,             
                  Universite de Grenoble 1, Grenoble, France}                 
\centerline{$^{8}$Centre de Physique des Particules de Marseille,             
                  IN2P3-CNRS, Marseille, France}                              
\centerline{$^{9}$LPNHE, Universit\'es Paris VI and VII, IN2P3-CNRS,          
                  Paris, France}                                              
\centerline{$^{10}$DAPNIA/Service de Physique des Particules, CEA, Saclay,    
                  France}                                                     
\centerline{$^{11}$Panjab University, Chandigarh, India}                      
\centerline{$^{12}$Delhi University, Delhi, India}                            
\centerline{$^{13}$Tata Institute of Fundamental Research, Mumbai, India}     
\centerline{$^{14}$Seoul National University, Seoul, Korea}                   
\centerline{$^{15}$CINVESTAV, Mexico City, Mexico}                            
\centerline{$^{16}$Institute of Nuclear Physics, Krak\'ow, Poland}            
\centerline{$^{17}$Institute for Theoretical and Experimental Physics,        
                   Moscow, Russia}                                            
\centerline{$^{18}$Moscow State University, Moscow, Russia}                   
\centerline{$^{19}$Institute for High Energy Physics, Protvino, Russia}       
\centerline{$^{20}$Lancaster University, Lancaster, United Kingdom}           
\centerline{$^{21}$University of Arizona, Tucson, Arizona 85721}              
\centerline{$^{22}$Lawrence Berkeley National Laboratory and University of    
                   California, Berkeley, California 94720}                    
\centerline{$^{23}$University of California, Davis, California 95616}         
\centerline{$^{24}$California State University, Fresno, California 93740}     
\centerline{$^{25}$University of California, Irvine, California 92697}        
\centerline{$^{26}$University of California, Riverside, California 92521}     
\centerline{$^{27}$Florida State University, Tallahassee, Florida 32306}      
\centerline{$^{28}$University of Hawaii, Honolulu, Hawaii 96822}              
\centerline{$^{29}$Fermi National Accelerator Laboratory, Batavia,            
                   Illinois 60510}                                            
\centerline{$^{30}$University of Illinois at Chicago, Chicago,                
                   Illinois 60607}                                            
\centerline{$^{31}$Northern Illinois University, DeKalb, Illinois 60115}      
\centerline{$^{32}$Northwestern University, Evanston, Illinois 60208}         
\centerline{$^{33}$Indiana University, Bloomington, Indiana 47405}            
\centerline{$^{34}$University of Notre Dame, Notre Dame, Indiana 46556}       
\centerline{$^{35}$Purdue University, West Lafayette, Indiana 47907}          
\centerline{$^{36}$Iowa State University, Ames, Iowa 50011}                   
\centerline{$^{37}$University of Kansas, Lawrence, Kansas 66045}              
\centerline{$^{38}$Kansas State University, Manhattan, Kansas 66506}          
\centerline{$^{39}$Louisiana Tech University, Ruston, Louisiana 71272}        
\centerline{$^{40}$University of Maryland, College Park, Maryland 20742}      
\centerline{$^{41}$Boston University, Boston, Massachusetts 02215}            
\centerline{$^{42}$Northeastern University, Boston, Massachusetts 02115}      
\centerline{$^{43}$University of Michigan, Ann Arbor, Michigan 48109}         
\centerline{$^{44}$Michigan State University, East Lansing, Michigan 48824}   
\centerline{$^{45}$University of Nebraska, Lincoln, Nebraska 68588}           
\centerline{$^{46}$Columbia University, New York, New York 10027}             
\centerline{$^{47}$New York University, New York, New York 10003}             
\centerline{$^{48}$University of Rochester, Rochester, New York 14627}        
\centerline{$^{49}$State University of New York, Stony Brook,                 
                   New York 11794}                                            
\centerline{$^{50}$Brookhaven National Laboratory, Upton, New York 11973}     
\centerline{$^{51}$Langston University, Langston, Oklahoma 73050}             
\centerline{$^{52}$University of Oklahoma, Norman, Oklahoma 73019}            
\centerline{$^{53}$Brown University, Providence, Rhode Island 02912}          
\centerline{$^{54}$University of Texas, Arlington, Texas 76019}               
\centerline{$^{55}$Texas A\&M University, College Station, Texas 77843}       
\centerline{$^{56}$Rice University, Houston, Texas 77005}                     
\centerline{$^{57}$University of Washington, Seattle, Washington 98195}       
}                                                                             

\maketitle

\begin{abstract}
We report on a measurement of 
$\sigma({\mbox{$p\overline{p} \rightarrow W+X$}}) \cdot 
B({\mbox{$W \rightarrow \tau \nu$}})$ in $p\overline{p}$ collisions
at $\sqrt{s} = 1.8$ TeV at the Fermilab Tevatron.
The measurement is based on an integrated luminosity of 18 pb$^{-1}$
of data collected with the D\O\ detector during 1994--1995.
We find that
${\sigma({\mbox{$p\overline{p} \rightarrow W+X$}}) \cdot B({\mbox{$W
\rightarrow \tau \nu$}}) = 2.22 \pm 0.09~{\rm (stat)} \pm 0.10~{\rm (syst)}
\pm 0.10~{\rm (lum)}} {\rm ~nb.} \nonumber
$
Lepton universality predicts that the ratio of the tau and electron electroweak
charged current couplings to the $W$ boson, $g_\tau^W / g_e^W$, 
be unity.
We find
$
g_\tau^W / g_e^W = 0.980 \pm 0.031,
$
in agreement with lepton universality.
\end{abstract}



The measurement of the $W$ boson production cross section times branching
ratio to  $\tau$ lepton and neutrino, $\sigma(p\overline{p}\rightarrow W+X) 
\cdot
B(W \rightarrow \tau\nu)$, can be used with the corresponding result
from the electron channel, $\sigma(p\overline{p}\rightarrow W+X) \cdot
B(W \rightarrow e\nu)$, to test one of the fundamental concepts in the 
standard model:
the universality of the leptonic couplings to the weak charged current.  Such
``lepton universality" is
a direct consequence of SU(2) gauge symmetry
and the assumption that the leptons transform as left-handed SU(2) doublets,
making its characterization a basic test of the underlying structure
of the theory.
Previous tests of $\tau$-$e$ universality at high $Q^{2}$ ($Q^2 \approx M_W^2$)
have been obtained from
the direct measurements of $\sigma_W\cdot B(W\rightarrow \tau\nu$) and
$\sigma_W\cdot B(W\rightarrow e\nu$)
by the UA1\cite{ua1},
UA2\cite{ua2} and CDF\cite{cdf}
collaborations. 
Results from the CERN $e^+e^-$ collider (LEP) on the couplings of $Z$
bosons to charged leptons support three-generation lepton universality
to a precision of 0.5\%~\cite{pdg}.  Recent measurements of 
$B(W \rightarrow \tau
\nu)$ from $WW$ production at LEP~\cite{LEPB} are consistent with
lepton universality, and low $Q^2$ measurements of $\tau$-lepton decay
branching fractions~\cite{pdg} also support lepton universality.

In this Letter we report a new measurement of
$\sigma({\mbox{$p\overline{p} \rightarrow W+X$}}) \cdot B({\mbox{$W
\rightarrow \tau \nu$}})$ using data collected with the D\O\ detector
during the 1994--1995  Fermilab Tevatron collider run
at a $p\overline{p}$ center-of-mass energy of $\sqrt{s} = 1.8$ TeV.
The integrated luminosity~\cite{wenu} for the
$\tau$ trigger used for this measurement is ${\mbox{$\int {\cal{L}} dt$}} = 
18.04
\pm 0.79$ pb$^{-1}$.  The D\O\ detector is described in detail in
Ref.~\cite{nim}.  The detector consists of a non-magnetic tracking system,
a uranium/liquid-argon calorimeter with segmentation $\Delta\eta
\times\Delta\phi = 0.1\times0.1$ in pseudorapidity and azimuth, and an 
iron toroid muon spectrometer.

In D\O\ the $\tau$ lepton is identified through its hadronic decay modes into
final states consisting of one or three charged hadrons plus neutral particles.
The $\tau$ decay products are highly boosted, forming a very narrow hadronic
jet. The signature for $W \rightarrow \tau \nu$, with $\tau \rightarrow 
\nu +$ hadrons, is therefore an isolated and very narrow
hadronic jet with low charged particle multiplicity, accompanied by a large
amount of missing transverse energy {\mbox{$\not\!\!E_T$}}, 
determined from the energy deposition in the calorimeter within $|\eta|<4.5$.

The $\tau$ trigger requires {\mbox{$\not\!\!E_T$}}$ > 16$ GeV, a leading 
(highest $E_T$)
narrow jet with transverse energy 
$E_T > 20 $ GeV and $ 0.05 < f_{\rm EM} < 0.95$,
where $f_{\rm EM}$
is the fraction of the jet energy in the electromagnetic
calorimeter.
The trigger also requires no jet with
$E_T > 15 $ GeV within $0.7$ radians in $\phi$ of the direction opposite to 
that of the leading jet,
or within $0.5$ radians in $\phi$ of the {\mbox{$\not\!\!E_T$}} direction,
where $\phi$ is the azimuthal angle.
In addition, a single interaction requirement is applied at the trigger level.

\begin{figure} [b] 
\epsfxsize=3.375in
\centerline {\epsfbox{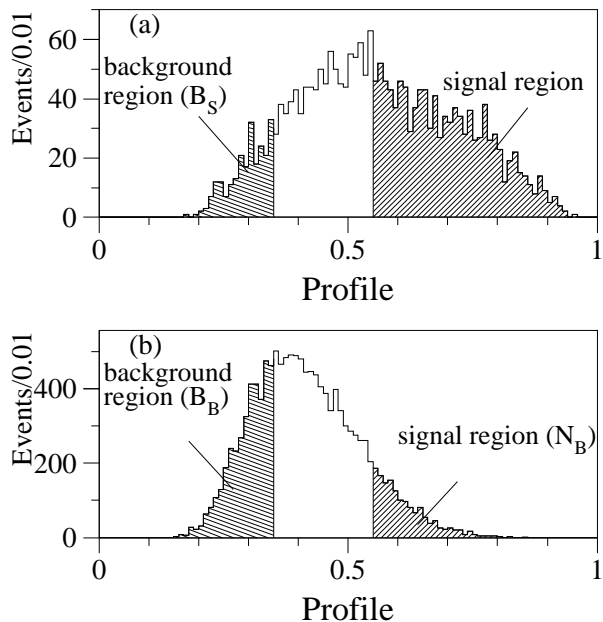}}
\vspace{0.2cm}
\caption{The ${\cal P}$ distributions of (a) the $\tau$ sample from data, 
and (b) the QCD
background sample.}
\label{fig:profile_data_qcd}
\end{figure}

In the offline analysis, jets are reconstructed using a cone algorithm with
radius ${\cal R} = 0.7$ in $\eta$-$\phi$ space, where
$\eta$ is the pseudorapidity.  $W \rightarrow \tau \nu$ events
are selected by requiring one jet satisfying 
({\it i}) $25 < E_T < 60$ GeV,
({\it ii}) jet width ${\cal W} \leq 0.25$, where  
$$
 {\cal W} = \sqrt{\sum_{i=1}^{n}
\frac{{\Delta\phi}^2 E_{Ti}}{E_T} + \sum_{i=1}^{n}
\frac{ {\Delta\eta}^2 E_{Ti} }{E_T} }
$$
and $i=1, \ldots, n$ indicates the calorimeter $\eta$-$\phi$ tower number,
({\it iii}) $ 0.10 < f_{\rm EM} < 0.95$,
({\it iv}) $|\eta| \leq 0.9$,
({\it v}) one to seven reconstructed tracks within a $0.2 \times 0.2$ road
         in $\eta$-$\phi$ space around the jet axis,
({\it vi}) at least one track within 0.1 radian in $\phi$ of the
          center of gravity of the jet,
({\it vii}) jet quality cuts involving the longitudinal and lateral distribution
           of the energy within the jet, and
({\it viii}) profile ${\cal P} \geq 0.55$, where
$$ {\cal P} = \ {(E_{T1} + E_{T2}) \ / \  E_T}, $$
\vskip -0.2cm
\noindent
and $E_{T}, E_{T1}$ and $E_{T2}$ are the transverse energy of the jet and 
the two towers within the jet with the largest $E_T$, respectively.
The profile variable
exploits the fine calorimeter segmentation and good
energy resolution of the D\O\ detector. 
The very narrow jets from hadronic $\tau$ decays 
lead to high values of ${\cal P}$. 
QCD processes yield events with wider jets, 
and therefore lower values of ${\cal P}$ (Fig.~\ref{fig:profile_data_qcd}).

In addition, the event must have
({\it i}) {\mbox{$\not\!\!E_T$}}$> 25$ GeV,
({\it ii}) a $z$ vertex position within 60 cm of the detector center,
({\it iii}) no electrons or muons with $E_T > 15$ GeV,
({\it iv}) no jets with $E_T \geq 8$ GeV within $0.5$ radians of the
          {\mbox{$\not\!\!E_T$}} direction,
({\it v}) no jets with $E_T \geq 8$ GeV within 0.7 radians in $\phi$ of
         the direction opposite to that of the $\tau$ jet, and
({\it vi}) no jet with $E_T > 15$ GeV in addition to the $\tau$ jet.

The $\tau$ lepton identification is very sensitive to electronic noise in the
calorimeter and to the underlying event.
A data-based Monte Carlo (DBMC), using
{\mbox{$W \rightarrow e \nu$}} data, was developed to model
$W \rightarrow \tau \nu$ events with actual noise and underlying event effects.
We replace the electron from $W$ boson decays with a Monte Carlo $\tau$,
which was generated with the same kinematics as the electron, forced to
decay hadronically, and then passed through a detector simulation based on the
{\sc geant} Monte Carlo program~\cite{geant}.  
The tracking hits along the electron track and the 
        calorimeter cells associated with the electron cluster are replaced
        by the simulated Monte Carlo $\tau$ information.
In this way, only the $\tau$
decays and the response of the detector to the $\tau$ decay products are
simulated with a Monte Carlo, and noise and underlying event effects are taken
directly from the data. 

The dominant background in the $W \rightarrow \tau \nu$ final sample is from
multijet events, in which one of the jets mimics a $\tau$ jet, and the
energies of the other jets fluctuate to give {\mbox{$\not\!\!E_T$}}.
We estimate this QCD background 
using the ${\cal P}$ distribution. The cuts to select the
``QCD background sample" are  similar to those used
to select the
 $W \rightarrow \tau \nu$ sample, but without the ${\cal P}$ 
or {\mbox{$\not\!\!E_T$}} requirements or the requirement  that there be no 
jet with $E_T > 15$ GeV in addition to the leading jet.  
The ``$\tau$ sample" is the final
$W \rightarrow \tau \nu$ sample before the ${\cal P}$ cut.  We define the
region with ${\cal P} < 0.35$ as the ``background region," and the region with
${\cal P} > 0.55$ as the ``signal region," as shown in
Fig.~\ref{fig:profile_data_qcd} for both the $\tau$ sample and the QCD 
background sample. 
We find that the ${\cal P}$ distribution of the background sample is 
uncorrelated with {\mbox{$\not\!\!E_T$}}, leading jet $E_T$ or the 
number of jets in the event.  
From the DBMC $W \rightarrow \tau \nu$ studies only $\sim 1$\% of
$W \rightarrow \tau \nu$ events are in the background region.
The number of background events in the signal region of the $\tau$ sample
($N_{\rm QCD}$) can be calculated as
$$
N_{\rm QCD} = N_B \times (B_S / B_B),
$$
where $N_B=1834$ 
is the number of events in the signal region of the QCD background
sample, $B_B=4422$ is the number of events in the background region of the QCD
background sample, and $B_S=253$ is
the number of events in the background region of
the $\tau$ sample. 
We obtain
$N_{\rm QCD} = 106\pm7~{\rm(stat)}\pm5~{\rm(syst)}$ events \cite{lihl}.  The systematic
error is estimated from the dependence of the average 
profile on {\mbox{$\not\!\!E_T$}} in the background 
region of the QCD background sample.
We compared the ${\cal P}$ distribution in the
background region between the QCD background sample and the $\tau$ sample. 
Their shapes agree very well -- the Kolmogorov-Smirnov probability that the two
distributions are from the same parent distribution is 0.94. This   
assures us of the validity of normalizing the
background region in the QCD background sample to the background region in the
$\tau$ sample. As another consistency check, we divided the $\tau$
 and QCD background
samples into bins in {$E_T$} of the $\tau$ jet 
and calculated $N_{\rm QCD}$ separately
for each {$E_T$} bin. We estimated $N_{\rm QCD}$ to be 107 events
using this method. 
We also checked the ${\cal P}$ distribution in 
{\mbox{$\not\!\!E_T$}} bins. We see no significant dependence
of ${\cal P}$ on {\mbox{$\not\!\!E_T$}}.

The QCD background estimate also includes $W/Z+$jet events 
in which a jet is misidentified as a $\tau$ jet and the {\mbox{$\not\!\!E_T$}} 
arises from either a $W$ leptonic decay or from unreconstructed muons in 
{\mbox{$Z \rightarrow \mu\mu$} decays,  
which result in {\mbox{$\not\!\!E_T$}} in the calorimeter.
The background from $Z+$jet events in which {\mbox{$Z \rightarrow \nu\nu$} is 
also included in the QCD background estimate. 
The {\mbox{$W \rightarrow e \nu$}} background, in which the electron is
misidentified as a $\tau$, is estimated to be $3 \pm 1$ events.

Electronic noise in a calorimeter cell may simulate a narrow jet and
also give a large {\mbox{$\not\!\!E_T$}}. When an underlying event track 
is very close to the noise jet, it mimics a $\tau$ event.
The background from noise events is estimated by using
the same method that was used to calculate $N_{\rm QCD}$, but  
using the distribution of $\Delta \phi$, the difference in $\phi$ of the 
$\tau$ jet and the closest track, instead of the ${\cal P}$ distribution. 
We defined $\Delta \phi > 0.1$ as the background region.  This gives
the number of background noise events in the final $\tau$ sample to be
$81\pm14$.  

\begin{table}[tb]
\label{all}
\caption{Summary of the $\sigma({\mbox{$p\overline{p} \rightarrow W+X$}})\
\cdot B({\mbox{$W \rightarrow \tau \nu$}})$ measurement.}
\vskip 0.15 cm
\begin{tabular}{l|c}
\hspace{.2 cm}
$N_{\rm obs}$                      & $ 1202  $ \\
\hline
\hspace{.2 cm}
Backgrounds (No. events):          &           \\
\hspace{.5 cm}
QCD $\, \pm$(stat)$\pm$(syst)      & $ 106 \pm  7 \pm 5 $ \\
\hspace{.5 cm}
Electronic noise                   & $ 81  \pm 14 $ \\
\hspace{.5 cm}
$Z \rightarrow \tau \tau$          & $ 32  \pm  5 $ \\
\hspace{.5 cm}
$W \rightarrow e \nu$              & $  3  \pm  1 $ \\
\hspace{.2 cm}
Total Background (No. events)      & $ 222 \pm 17 $ \\
\hspace{.2 cm}
$A \cdot \epsilon $                 & $ 0.0379 \pm 0.0017 $ \\
\hline
\hspace{.2 cm}
\mbox{\normalsize$\int {\cal{L}}dt\ ({\mbox{\rule[0cm]{0cm}{0.4cm}${\rm pb}^{-1}$}})$}  & $ 18.04 \pm 0.79 $\\
\hline
\hspace{.2 cm}
{${\sigma_W\cdot} B(W \rightarrow \tau \nu)$\enskip (nb)}  &   \\
\hspace{.6 cm}$\pm$ (stat),(syst),(lum)   & $  2.22 \pm 0.09 \pm 0.10 \pm 0.10
$ \hspace{.3 cm} \\
\end{tabular}
\label{tbl:summary}
\end{table}

\begin{figure} [b]
\epsfxsize=3.375in
\centerline {\epsfbox{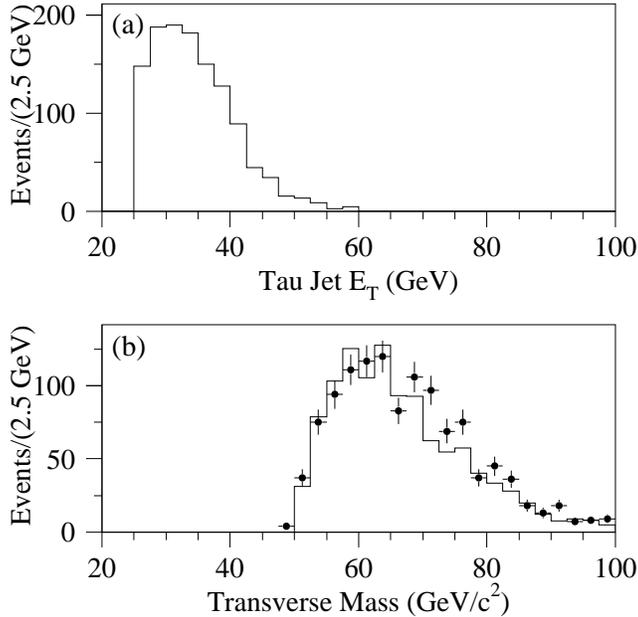}}
\vspace{0.2cm}
\caption{
(a) The $\tau$ jet $E_T$ distribution for data passing all the selection 
cuts, and 
(b) the distribution of the transverse mass of the $\tau$ jet and the 
{\mbox{$\not\!\!E_T$}} for DBMC events 
(histogram) and for data (points) passing
all the selection cuts and with QCD background subtraction.}
\label{fig:et_mt}
\end{figure}

Another source of background is {\mbox{$Z \rightarrow \tau \tau$}},
where one of the $\tau$ leptons decays hadronically.  We studied this 
background using
the {\sc isajet}~\cite{isajet} generator and the {\sc geant}-based D\O\
simulation program. Applying the same cuts 
as those used in the  $W \rightarrow \tau \nu$ event selection, 
we estimate that $32 \pm 5$
{\mbox{$Z \rightarrow \tau \tau$}} events are present in our final data sample.
The number of background events is summarized in
Table~\ref{tbl:summary}.

There are 1202 events passing all the selection cuts.  For these events
Fig.~\ref{fig:et_mt}(a) shows the $\tau$ jet $E_T$ distribution.
Figure~\ref{fig:et_mt}(b) shows the distribution of the transverse mass
calculated from the $\tau$ jet and the {\mbox{$\not\!\!E_T$}} for the
$\tau$ sample after QCD background subtraction. 
Figure~\ref{fig:et_mt}(b) also shows the comparison with DBMC events,
normalized to the data, passing the same cuts.
The distribution of jet width (${\cal W}$) can also be used to confirm the 
selection
of $W \rightarrow \tau \nu$ events.  Figure~\ref{fig:WIDTH} compares the
${\cal W}$ distribution for DBMC $\tau$ jets and QCD jets  before and after
the profile (${\cal P}$) cut.  Figure~\ref{fig:WIDTH}(b) also shows the
${\cal W}$ distribution for $\tau$ jets from the final data sample with the QCD
background subtracted.
The
${\cal W}$ distribution of the final data sample is clearly different 
from that of 
the QCD jet sample, and agrees well with the DBMC $W \rightarrow \tau \nu$
prediction.

\begin{figure} [b]
\epsfxsize=3.375in
\centerline {\epsfbox{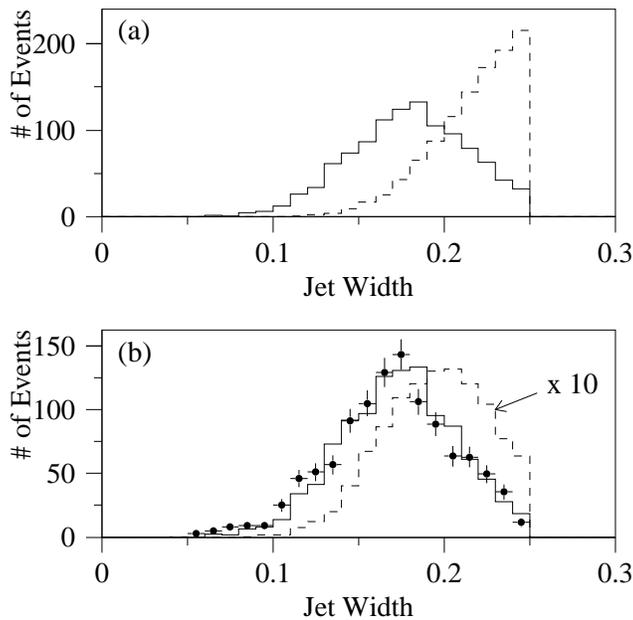}}
\caption{
The distribution of jet width for DBMC $\tau$ jets (solid histogram)
and QCD jets (dashed histogram) (a) before the profile cut (with
arbitrary scale), and (b) after
the profile cut. The  ${\cal W}$ distribution for data (points) 
is also shown in (b). The QCD distribution in (b) has been scaled up by a 
factor of 10.}
\label{fig:WIDTH}
\end{figure}

The acceptance $A$ is determined by applying the geometric and kinematic cuts
on {\sc isajet} Monte Carlo $\tau$ leptons, giving $ A = 0.2903 \pm 0.0007$.
The efficiency $\epsilon$ is determined by applying the trigger requirements 
and the offline cuts on the DBMC $W \rightarrow \tau \nu$ sample, giving
$\epsilon = 0.1307\pm0.0034$.
The trigger efficiency for the events passing the offline
selection is $0.9941\pm0.0020$.
The above uncertainties are from Monte Carlo statistics and are treated as 
systematic.  
Two additional sources contribute to systematic uncertainties in
$A \cdot \epsilon$.  
First, the 3\% uncertainty in the energy
scale~\cite{cafix} results in an uncertainty of 2.8\%
on $A \cdot \epsilon$.
Second, the uncertainty due to $\tau$ branching fraction uncertainties 
is calculated by varying the branching fractions of the various 
exclusive $\tau$ decay modes by their
measurement errors~\cite{pdg}, subject to the constraint that the sum of all
$\tau$ decay branching fractions add up to 1. This variation results in an uncertainty of
2.0\%
on $A \cdot \epsilon$.
The final value of $A \cdot \epsilon$ is $0.0379 \pm 0.0017$.

The cross section times branching ratio for 
$p\overline{p}\rightarrow W+X$, with
$W \rightarrow \tau \nu$, is calculated using the formula
$$
\sigma_W \cdot B(W \rightarrow \tau \nu) = \frac{N_{\rm obs} - N_{\rm bkg}}{
{\int {\cal{L}} dt} \cdot
B({\tau \rightarrow \nu + \rm {hadrons}}) \cdot A \cdot \epsilon },
$$
where $N_{\rm obs}$ is the number of events in the final data sample, $N_{\rm bkg}$ is
the estimated background, $A$ is the acceptance, $\epsilon$ is the efficiency,
${\int {\cal{L}} dt}$ is the integrated luminosity, and
$B({\tau \rightarrow \nu + \rm {hadrons} }) = (64.69\pm0.22)\% $~\cite{pdg}.
We measure
$$
\sigma_W \cdot B(W \rightarrow \tau \nu) =
2.22 \pm 0.09 \pm 0.10 \pm 0.10~{\rm nb},
$$
where the uncertainties are statistical, systematic, and due to the luminosity 
uncertainty, respectively.

We can determine the ratio of the tau and electron electroweak charged current
couplings to the $W$ boson, $g_\tau^W$ and $g_e^W$, from
\begin{equation}
\label{eq:ratio}
{\left( \frac{g_\tau^W}{g_e^W} \right) }^2 =
\frac{\sigma(p\overline{p}\rightarrow W+X)\cdot B(W \rightarrow \tau\nu)}
{\sigma(p\overline{p}\rightarrow W+X)\cdot B(W \rightarrow e\nu)}.
\end{equation}
Taking the ratio of $\sigma_W \cdot B({\mbox{$W \rightarrow \tau \nu$}})$
and $\sigma_W \cdot B({\mbox{$W \rightarrow e \nu$}})$
completely cancels 
the luminosity error. 
Using our measurement~\cite{wenu} of
$\sigma_W \cdot B({\mbox{$W \rightarrow e \nu$}}) = 2.31 \pm 0.01 \pm 0.05
\pm 0.10~{\rm nb}$ for data collected during the same Tevatron 
collider run, we find
\begin{eqnarray*}
g_\tau^W / g_e^W &=& 0.980 \pm 0.020~{\rm (stat)} \pm 0.024~{\rm (syst)} \\
&=& 0.980 \pm 0.031. 
\end{eqnarray*}
\noindent
Phase space effects and non-universal radiative corrections will
modify Eq.~\ref{eq:ratio}, but the resulting uncertainties on $g_\tau^W / g_e^W$
are negligible compared with the experimental uncertainty in this
result. 

Our measurement is in good agreement with lepton universality,
which requires that $g_\tau^W / g_e^W = 1$.
Figure~\ref{fig:results} shows the results for $g_\tau^W / g_e^W$ from other
experiments, along with the value determined by the D\O\ experiment and the
weighted average of the four experiments, which is $0.988 \pm 0.025$.
The average was calculated assuming systematic errors are
uncorrelated among the four experiments.

In summary, we have used the D\O\ detector to identify $\tau$ leptons in
$p\overline{p}$ collisions, have measured the cross section times branching
ratio $\sigma({\mbox{$p\overline{p} \rightarrow W+X$}})\
\cdot B({\mbox{$W \rightarrow \tau \nu$}})$, and have used this result to
test $\tau$-$e$ universality at high $Q^{2}$ ($Q^2 \approx M_W^2$) to a 
precision of 3\%.

\begin{figure}[t]
\epsfxsize=3.375in
\centerline {\epsfbox{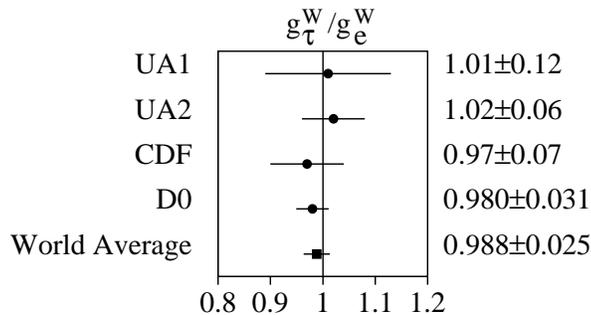}}
\caption{
$g_\tau^W / g_e^W$ from UA1~[1], UA2~[2], CDF~[3], and this measurement (D\O).}
\label{fig:results}
\end{figure}

%
We thank the Fermilab and collaborating institution staffs for 
contributions to this work, and acknowledge support from the 
Department of Energy and National Science Foundation (USA),  
Commissariat  \` a L'Energie Atomique (France), 
Ministry for Science and Technology and Ministry for Atomic 
   Energy (Russia),
CAPES and CNPq (Brazil),
Departments of Atomic Energy and Science and Education (India),
Colciencias (Colombia),
CONACyT (Mexico),
Ministry of Education and KOSEF (Korea),
CONICET and UBACyT (Argentina),
A.P. Sloan Foundation,
and the Humboldt Foundation.

\end{document}